# A Review of Automated Diagnosis of COVID-19 Based on Scanning Images

Delong Chen[1,2], Shunhui Ji[2] [*], Fan Liu[1,2], Zewen Li[2], Xinyu Zhou[3]

[1] Key Laboratory of Ministry of Education for Coastal Disaster and Protection, Hohai University, Nanjing, China
[2] College of Computer and Information, Hohai University, Nanjing, China
[3] China Pharmaceutical University, Nanjing, China
*shunhuiji@hhu.edu.cn

## ABSTRACT

The pandemic of COVID-19 has caused millions of infections, which has led to a great loss all over the world, socially and economically. Due to the false-negative rate and the time-consuming of the conventional Reverse Transcription Polymerase Chain Reaction (RT-PCR) tests, diagnosing based on X-ray images and Computed Tomography (CT) images has been widely adopted. Therefore, researchers of the computer vision area have developed many automatic diagnosing models based on machine learning or deep learning to assist the radiologists and improve the diagnosing accuracy. In this paper, we present a review of these recently emerging automatic diagnosing models. 70 models proposed from February 14, 2020, to July 21, 2020, are involved. We analyzed the models from the perspective of preprocessing, feature extraction, classification, and evaluation. Based on the limitation of existing models, we pointed out that domain adaption in transfer learning and interpretability promotion would be the possible future directions.

## CCS Concepts

• **Computing methodologies → Artificial intelligence → Computer vision→Computer vision tasks→Biometrics.**

## Keywords

Deep learning, Machine learning, Biomedical Image Analysis, COVID-19.

## 1. INTRODUCTION

It has been seven months since the first case of COVID-19 was confirmed. In the battle between human and the novel coronavirus, early diagnosing and early quarantine is of vital importance. However, testing based on Reverse Transcription Polymerase Chain Reaction (RT-PCR) is time-consuming and may cause certain false-negative reports. To solve this problem, diagnosing based on scanning images (CT or X-ray) has been proved to be practical and effective. In the virus-stricken area, radiologists have a heavy burden on analyzing scanning images. As shown in Fig. 1, researchers therefore have started to pay more and more attention to the development of COVID-19 diagnosing models for reducing the diagnosing time and improve the accuracy of radiologists.

Due to the rapid development in this area, there have already been 9 reviews ([79]-[87]) existing on this topic, but they have various shortcomings. To our best knowledge, there are at least 70 deep learning based and machine learning based models that have been proposed, and many of them have not been covered by any of the existing surveys. Most reviews only covered about 10 different diagnosing models. Moreover, these reviews lack proper organization, comparison of performance and in-depth analysis of shortcomings of diagnosing models.

Therefore, in this paper we define a universal pipeline for diagnosing models, for both machine learning based models and deep learning based models. Then we organized the paper according to different stages of the model. The contributions of this paper are as follows:

➢ We systematically reviewed and analyzed 70 COVID-19 diagnosing models from the perspective of preprocessing, feature extraction, classification, and evaluation. These models are proposed from February 14 to July 21, 2020.

➢ Based on the discussion of the existing models' limitation, we pointed out that domain adaption in transfer learning and interpretability promotion are the possible future directions.

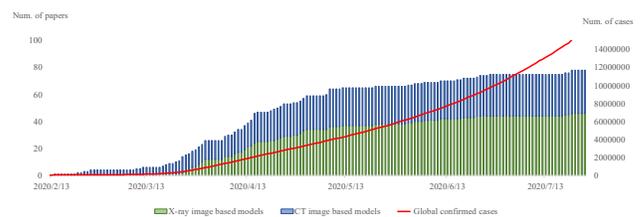

**Fig 1. Research of COVID-19 diagnosing models.**

## 2. DIAGNOSIS METHODS OF COVID-19

Diagnosing COVID-19 based on scanning images is regarded as a classification task by most researchers. These classification models share a similar pipeline. First, the lung scanning images (CT or X-ray) are preprocessed by data augmentation or lung segmentation, then the feature vectors are extracted by Convolution Neural Networks (CNN) or other feature extractors. The classifier predicts the infection. Finally, the model output heat map or bounding box to interpret its diagnosing result. In the following sections, we will review methods adopted in each stages of diagnosing models.

### 2.1 Preprocessing

In the existing literature, researchers have proposed three types of preprocessing methods: augmentation, equalization and segmentation. Data augmentation can enlarge the dataset and prevent overfitting, equalization improves the image quality, while lung segmentation can preserve the region of interest (ROI) only and avoid the undesired interfere from areas out of the lung.

---

[*] Corresponding author

To avoid overfitting, data augmentations such as rotating, flipping, scaling, cropping, brightness and contrast adjusting are commonly adopted in the preprocessing stage [5,6,11,19,26,29,30,32,37,43,46,50,64,65,71], but their augment abilities are limited. In the comparative experiment of Mizuho et al. [67], the conventional data augmentation method improved the diagnosing performance by only 4%. Therefore, researchers proposed to apply other advanced data augmentation methods. Nour et al. [25] and Arvan et al. [40] respectively used Generative Adversarial Network (GAN) and Conditional Generative Adversarial Network (CGAN) to generate virtual samples for training. Generative models can significantly increase the dataset size, but the quality of the generated sample is difficult to guarantee.

To reduce the interference caused by different scanners and enhance the image contrast, Md et al. [31] and Oh et al. [34] performed histogram equalization on the images. But histogram equalization has potential harm of affecting image details or bringing unexpected noise. Md et al. [31] eliminated the noise by introducing Perona-Malik filter (PMF), while Lv et al. [52]and Manu et al. [72] solved the problem by proposing Contrast Limited Adaptive Histogram Equalization (CLAHE).

In both x-ray and CT images, areas out of the lung could interfere with the diagnosing model. Lung segmentation can reduce such undesired effects by preserving the region of interest (ROI) only. U-Net based methods were used in [5,6,8,28,38,41,51,52] for fast, automatic and accurate lung segmentation. For CT images, performing lung segmentation slice by slice will lose the contextual information between slices. Therefore, some researchers [2][11] applied 3D versions of U-Net for lung segmentation, such as V-Net and 3D U-Net++. Besides, other methods such as Dense-Net in [13][34], OpenCV in [3], DeepLab in [9], and NABLA-N [32] were also adopted for lung segmentation.

**Table I. CNNs Used By COVID-19 Diagnosing Models**

| CNN Structure | Paper | Total |
|---|---|---|
| ResNet | [2,3,5,7,8,9,11,12,16,17,18,19,20,24,25,28,31,33,35,42,43,44,45,48,50,52,65,67,68,75] | 32 |
| GoogLeNet | [1,7,11,12,14,20,25,37,41,42,44,45,48,68,69,75] | 16 |
| DenseNet | [3,12,19,31,36,41,43,44,45,50,51,52,54,67,68,75] | 16 |
| VGG | [3,12,14,31,33,35,41,44,48,67,68,70,73,76] | 14 |
| MobileNet | [12,14,27,36,41,44,50,67,75] | 9 |
| SqueezeNet | [7,19,25,43,46] | 6 |
| AlexNet | [7,15,19,23,25] | 5 |
| Capsule | [26,40] | 2 |

## 2.2 Feature Extraction

Feature extraction is to detect those discriminative lesion patterns such as Ground Glass Opacity (GGO) in scanning images. As summarized in Table I, most COVID-19 diagnosing models adopted existing Convolutional Neural Network (CNN) structure for feature extraction. Besides, some researchers also proposed automatic network structure designing methods to identify the best network structure or hyperparameters for lung feature extraction such as generative synthesis approach [22], Gravitational Search Algorithm (GSA) [30] and iteratively pruning strategy [41]. Furthermore, model ensemble can also promote the overall performance. Lawrence et al. [35] and Umut et al [29] performed model ensemble by voting and feature fusion. Md et al. [31] apply Softmax Class Posterior Averaging (SCPA) and Prediction Maximization (PM) for model ensemble, and Rodolfo et al. [39] combined seven traditional feature extraction models with Inception-v3 to obtain better results. Mahesh et al. [73] ensemble different CNNs by a stacked generalization approach to further improve the model performance. These models assumed that different sub-models learn nonlinear discriminative features and semantic image representation from images of different levels. Therefore, the combined model could be more robust and accurate.

At the beginning of the pandemic, trying existing CNN is fast and convenient. However, these networks are designed for general image classification tasks such as ImageNet challenge and might be not suitable enough for scanning images classification. Radiologists diagnose COVID-19 by finding distinguishing local patterns. Some researchers design local methods to extract more discriminative features. For example, Umut et al. [29] and Oh et al. [34] used the local patches to train the CNN feature extractor. But lung infectious areas may vary significantly in size, the local methods with fixed patch size is unable to extract features of the target with the larger size. Hu et al. [38] proposed multi-scale learning to overcome such deficiency. The network aggregated features from different layers to make the final decision. Similarly, Ying et al. [3] integrated ResNet50 with the Feature Pyramid Network (FPN) , which is a pyramidal hierarchy network structure for multi-scale feature extraction. Besides, the lesion of COVID-19 in the lung is a 3D object, slice-wise contextual information in CT images would be lost by conventional 2D feature extractor. Therefore, Zheng et al. [6] proposed a CNN structure with 3D convolution units to detect COVID-19 to solve this defect.

In practice, radiologists also need to consider information such as epidemiology and clinical manifestations for diagnosis. Therefore, some methods also combine auxiliary external information with visual features to improve the model. Wang et al. [13] combine clinical features including age, sex, and comorbidity with CNN features. Similarly, since infected area usually lies near the edge of the lung, Xu et al. additionally provide the distance-from-edge information [2] of the local patch to the network. Shi et al. [10] and Sun et al. [53] calculate human-designed features including using volume, infection lesion number, histogram distribution, surface area, and radionics information.

## 2.3 Classification

Classification is to present diagnosing prediction according to the extracted feature. Most existing COVID-19 diagnosing models used CNN as the feature extractor, and most of them use softmax as the classifier. Some researchers proposed improvements based on the CNN + softmax scheme. For example, Wang et al. [1] combined softmax, decision tree, and Adaboost algorithms. Zhang et al. [16] simultaneously performed softmax loss based classification and contrastive loss-based anomaly detection to make the final decision. However, these deep models are black-box and usually need large-scale training sets. In literatures [7,21,29], researchers developed a non-end-to-end models and taking Support Vector Machine (SVM) as the classifier. Comparative experiments of classification algorithms including SVM, logistic regression, k-Nearest Neighbors (k-NN), Multi-Layer Perception (MLP), decision tree, AdaBoost, random forest, LightGBM , and Bagging Classifier have been done in [10,39,44,53], but they reached different conclusions. We think it is caused by different datasets and different feature extractors.

A straightforward way of modeling the COVID-19 diagnosing task into a classification task is binary classifying the scanning images into COVID-19 class and normal class. This way was adopted by many models including [4,5,6,9,12,16,20,21,23,27,29,32,40,43,44,46,49,65,69,71]. But in practice, test images of other types of abnormal lung can be misclassified as COVID-19. As shown in Fig. 2, diagnosing COVID-19 is a fine-grained task, lung diseases that belong to the same subclass share similar patterns in scanning images, and have a chance to be misclassified. To solve this problem, researchers overcome the problem of misclassification mainly through two approaches: multi-class classification and multi-step classification.

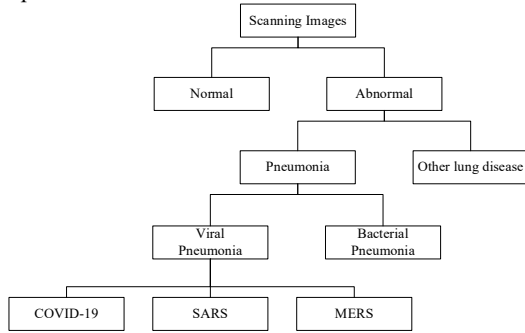

**Fig 2. Hierarchical relationships of lung disease**

For multi-class classification, some researchers added other common pneumonia categories in addition to the binary classification tasks, such as viral pneumonia, bacterial pneumonia, Community-Acquired Pneumonia (CAP), and non-COVID-19 pneumonia. This strategy is adopted by 27 different models. The classification task of COVID-19 / normal / viral pneumonia / bacterial pneumonia [17,22,24,26,31,35,37,64,68,77] and classification task of COVID-19 / normal / non-COVID-19 pneumonia [14,33,36,67,70,76]are the most popular setting. Some models also take account of other types of lung diseases, including ARDS [7][30], Tuberculosis [34], lung cancer [11][13], Pneumocystis, Streptococcus, Varicella [39], Fevers and upper respiratory tract symptoms [28].

However, multi-class classification methods rely heavily on datasets. Meanwhile, the model cannot learn the hierarchical relationships between categories. Multi-step classification can help the models to learn hierarchical relationships. For example, Eduardo et al. [33] and Yeh et al. [51] trained two binary classifiers, one for normal/pneumonia classification and one for further COVID-19 / non-COVID-19 classification. Lv et al. [52] firstly classify the screening image into normal / bacterial pneumonia / viral pneumonia, then perform the COVID-19 / non-COVID-19 classification. The above methods manually set up the hierarchical relationships for the models, while Rodolfo et al. [39] proposed to automatically learn a decision tree by the state-of-the-art Clus-HMC framework. They also comparatively tested multi-class classification and automatically multi-step classification, and reach a conclusion of the multi-step classification could be a feasible approach to improve COVID-19 recognition performance.

## 2.4 Evaluation

Researchers evaluated their proposed models with several metrics in experiments. The most used metrics are accuracy and the Area Under Curve (AUC). In Table II, we summarized the evaluation of the above-mentioned COVID-19 diagnosing models. The average accuracy and AUC of diagnosing models based on X-ray scanning are 94.76% and 96.94, and the average accuracy and AUC of CT scanning based models are 90.13% and 94.76. Theoretically, 3D CT scanning contains more information than 2D X-ray scanning, and CT scanning can also avoid the occlusion of ribs compared with X-ray scanning. However, X-ray scanning based models achieve better performance surprisingly. We consider the reason is the large size of X-ray training sets helps these models, while CT scanning is relatively more difficult to collect: the average training set size of X-ray based models is 4185, and the average size of CT based models is only 1417.

Although the performance of existing models is relatively high, but the size of the test set is worth noting. In some models, the test set has only a few COVID-19 samples. The average COVID-19 / total ratio of test sets is 0.274:1, which is highly imbalanced. In Table II we also color the table cells according to the number of samples in training and testing dataset. Some researchers reproduced the experiment with different dataset, but got significantly lower performance compared to the original reported performance [78]. The reason might be model overfitting and the lack of appropriate control patients and ground truth labels. Moreover, these models are evaluated in different datasets, and most of them are private datasets. We think a proper benchmark test set is vital for further research of this area.

**Table II. Dataset Size and Performance of COVID-19 Diagnosing Models**

| Date | Paper | Scanning type | | Training set | | Test set | | Model performance | |
|---|---|---|---|---|---|---|---|---|---|
| | | X-ray | CT | COVID-19 | Total | COVID-19 | Total | Accuracy | AUC |
| 14-Feb | [1] | - | √ | 44 | 99 | 119 | 237 | 73.10% | 78.00 |
| 21-Feb | [2] | - | √ | 219 | 618 | 30 | 90 | 86.70% | |
| 23-Feb | [3] | - | √ | 53 | 165 | 26 | 82 | 94.00% | 99.00 |
| 25-Feb | [4] | - | √ | 40 | 64 | 11 | 42 | 98.85% | |
| 10-Mar | [5] | - | √ | 108 | 243 | 12 | 27 | | 99.60 |
| 12-Mar | [6] | - | √ | 289 | 499 | 76 | 131 | 90.10% | 95.90 |
| 18-Mar | [7] | √ | - | 80 | 160 | 27 | 54 | 95.30% | |
| 19-Mar | [8] | - | √ | 400 | 3069 | 68 | 353 | | 96.00 |
| 20-Mar | [9] | - | √ | 222 | 312 | 183 | 1072 | 94.98% | 97.91 |
| 22-Mar | [10] | - | √ | 1326 | 2292 | 332 | 573 | 87.90% | 94.20 |
| 23-Mar | [11] | - | √ | 723 | 1136 | 154 | 282 | | 99.10 |
| 24-Mar | [12] | √ | - | 20 | 40 | 5 | 10 | 83.00% | 90.00 |
| 24-Mar | [13] | - | √ | 1266 | 5372 | 102 | 226 | 85.00% | 88.00 |
| 25-Mar | [14] | √ | - | 202 | 1284 | 22 | 143 | 97.82% | |
| 26-Mar | [15] | √ | - | 74 | 137 | 32 | 59 | 95.12% | 94.15 |
| 27-Mar | [16] | √ | - | 70 | 100 | 50 | 764 | 96.00% | 95.18 |
| 27-Mar | [17] | √ | - | 54 | 4753 | 14 | 1188 | 89.82% | |
| 28-Mar | [18] | - | √ | 100 | 513 | 50 | 256 | 98.80% | |
| 29-Mar | [19] | √ | - | 130 | 2876 | 60 | 180 | 98.30% | 99.80 |
| 29-Mar | [20] | √ | - | 40 | 80 | 10 | 20 | 98.00% | |
| 30-Mar | [21] | √ | - | 25 | 40 | unclear | unclear | 97.48% | |
| 30-Mar | [22] | √ | - | 76 | 16756 | 10 | 210 | 92.40% | |
| 31-Mar | [23] | √ | - | 76 | 16756 | 10 | 210 | 96.23% | |
| 31-Mar | [24] | √ | √ | 144 | 263 | 144 | 263 | 98.00% | |
| 2-Apr | [25] | √ | - | unclear | 624 | unclear | 1173 | 99.30% | |
| 6-Apr | [26] | √ | - | 70 | 100 | 50 | 764 | 95.70% | 97.00 |
| 6-Apr | [27] | √ | - | 409 | 3514 | 46 | 391 | 99.18% | |
| 6-Apr | [28] | - | √ | 829 | 1865 | 109 | 199 | | 99.40 |
| 7-Apr | [29] | - | √ | 2250 | 4500 | 750 | 1500 | 98.27% | |
| 9-Apr | [30] | √ | - | 70 | 114 | 15 | 31 | 98.00% | |
| 9-Apr | [31] | √ | - | 181 | 11896 | 78 | 5099 | 92.60% | |
| 10-Apr | [32] | √ | √ | 3875 | 5216 | unclear | 45 | 98.78% | |
| 12-Apr | [33] | √ | - | 66 | 16546 | 10 | 210 | 92.80% | |
| 12-Apr | [34] | √ | - | 126 | 354 | 36 | 99 | 88.90% | |
| 13-Apr | [35] | √ | - | 122 | 410 | 33 | 242 | 94.40% | 96.90 |
| 13-Apr | [36] | √ | - | 125 | 375 | 36 | 108 | | 96.50 |
| 14-Apr | [37] | √ | - | 256 | 1125 | 28 | 126 | 89.50% | |
| 14-Apr | [38] | - | √ | 120 | 360 | 30 | 90 | 89.20% | 92.30 |
| 15-Apr | [39] | √ | - | 63 | 802 | 300 | 342 | F1-score = 0.89 | |
| 15-Apr | [40] | √ | - | 286 | 14997 | 27 | 1703 | 99.01% | 99.72 |
| 15-Apr | [41] | - | √ | 286 | 625 | 47 | 105 | | 96.10 |
| 16-Apr | [42] | √ | - | 149 | 3783 | 31 | 11302 | 99.56% | |
| 20-Apr | [43] | √ | - | 31 | 2031 | 40 | 3040 | | 99.6 |
| 22-Apr | [44] | √ | - | 137 | 274 | unclear | unclear | 99% | |
| 23-Apr | [45] | √ | - | 88 | 881 | 11 | 108 | 98% | 99 |
| 24-Apr | [46] | √ | - | 180 | 718 | 45 | 175 | 95.30% | |
| 24-Apr | [47] | - | √ | 251 | 1768 | 108 | 203 | 83% | |
| 27-Apr | [48] | √ | - | 191 | 1791 | 48 | 448 | 97.01% | |
| 29-Apr | [49] | √ | - | 258 | 2799 | 60 | 945 | | 98.5 |

| Date | Ref | ✓ | ✓ | Col4 | Col5 | Col6 | Col7 | Acc | Extra |
|---|---|---|---|---|---|---|---|---|---|
| 30-Apr | [50] | √ | - | 175 | 12092 | 20 | 1509 | 89.40% | |
| 30-Apr | [51] | √ | - | 152 | 13594 | 31 | 231 | 96.80% | |
| 1-May | [52] | √ | - | 105 | 5486 | 10 | 405 | 83.12% | |
| 7-May | [53] | - | √ | 1196 | 2018 | 299 | 504 | 91.79% | 96.35 |
| 8-May | [54] | √ | - | 370 | 5029 | 92 | 1257 | 95.90% | |
| 6-May | [63] | - | √ | 1047 | 1765 | 449 | 757 | 93.90%% | |
| 8-May | [64] | √ | - | 370 | 5029 | 92 | 1257 | 95.90% | |
| 12-May | [65] | √ | - | 147 | 4086 | 37 | 1202 | 95.50% | |
| 22-May | [66] | √ | - | 101 | 131 | 152 | 197 | | 90.06 |
| 1-Jun | [67] | √ | - | 195 | 998 | 20 | 125 | 83.60% | |
| 3-Jun | [68] | √ | - | 189 | 989 | 47 | 247 | F1-score = 0.89 | |
| 6-Jun | [69] | √ | - | 1000 | 50000 | 565 | 22594 | | 94 |
| 11-Jun | [70] | √ | - | 415 | 8474 | 42 | 848 | 94.00% | |
| 16-Jun | [71] | - | √ | 314 | 671 | 35 | 75 | 86.80% | 3.1 |
| 18-Jun | [72] | √ | - | 429 | 1458 | 107 | 365 | 97.12% | |
| 22-Jun | [73] | √ | - | 189 | 1935 | 56 | 555 | 92.74% | |
| 23-Jun | [74] | - | √ | 657 | 3285 | 266 | 1330 | | |
| 26-Jun | [75] | - | √ | 35 | 75 | 315 | 675 | 90.61% | 96.05 |
| 19-Jul | [76] | √ | - | 398 | 5614 | 100 | 740 | 87.30% | |
| 21-Jul | [77] | √ | √ | 400 | 4685 | 50 | 1171 | 99.80% | |
| **Average** | - | - | - | 354 | 3804 | 96 | 1059 | 93.36% | 92.19 |

## 3. RESEARCH TREND ANALYSIS

### 3.1 Transfer Learning

A typical solution to the lack of massive scanning data is transfer learning. Among existing works, 34 models adopted the transfer learning scheme. They pre-trained the CNN on a larger image dataset (mostly on ImageNet), then fine-tune the model with X-ray or CT scanning images. But ImageNet contains images of general objects, which make the convolution filters to learn some patterns that will not appear in scanning images. Therefore, researchers proposed to transfer the model that pre-trained on lung cancer dataset [13] or conventional pneumonia dataset [51]. As another way to avoid overfitting, some researchers restricted [1,43,45, 47] or even skipped[7,27] the fine-tuning of the CNN feature extractor. However, these methods can reduce the chance of overfitting to some extent, but they did not unleash the full potential of deep models.

Domain adaptation is a branch of transfer learning, it's a learning technique to address the problem of lacking massive amounts of high-quality, large-scale labeled data. Fine-tuning only a certain part of the network can be regarded as domain adaptation [47], but there are also some specially designed deep domain adaptation models. Zhang et al. [49] used a domain discriminator to help the model better adapt to the target task. At present, there are few domain adaptation methods. we think applying deep domain adaptation to solve the problem of lacking massive training data of COVID-19 scanning images is an effective approach and valuable research direction.

### 3.2 Interpretability

In existing works, Gradient-weighted Class Activation Mapping (Grad-CAM) are adopted by researchers [5,8,9,16-18,22, 28,31,34,36,38,41,45,46,47,49,51,52] to output heatmaps for explaining the final result and present an intuitive understanding of which area is the model focusing on. At the same time, heatmaps can also provide radiologists with more useful information and further help them.

Compared to popular classification models, detection models can directly output a bounding box or a binary mask, they have an inherent advantage on interpretability. Detection based diagnosing is an emerging research direction. To the best of our knowledge, there are 5 detection based models[55]-[59] and 3 classification + detection models [60]-[62]. Radiologists search for lesions in the scanning images to make the diagnosis, so an object detection task can better simulate the human diagnosing process. Moreover, detection based methods can also avoid the information loss of local lesion patterns caused by the low dimension of the feature vector.

## 4. Conclusion

In this paper, we reviewed 70 automatic COVID-19 diagnosing models that emerged from February 14 to July 21, 2020. These models are based on machine learning or deep learning. They share a similar pipeline: preprocessing, feature extraction, classification, and evaluation. In the preprocessing stage of these models, transformation-based data augmentation and lung segmentation are performed. For feature extraction, most models adopted existing CNN structures, while others developed local methods to obtain more discriminative features. To enhance the performance of the classifiers, researchers proposed multi-class classification and multi-step classification. We also summarized the evaluation results of existing diagnosing models. Some of the models claim to perform well, but the size of test sets is not large enough.

Based on the limitations of existing models, we pointed out two possible future directions. Many existing models applied transfer learning to overcome the small dataset problem, but the adopted networks are pre-trained on general datasets such as ImageNet. To better utilize the information from both the source domain and target domain, a feasible solution is deep domain adaption. Besides, interpretability promotion is also important because it can further assist radiologists by providing more useful information. Detection based models have an inherent advantage on interpretability, such improvement for better interpretability is also a valuable direction.

## 5. ACKNOWLEDGMENTS

This work was partially funded by Natural Science Foundation of Jiangsu Province under grant No. BK20191298 and No. BK20170893, Fundamental Research Funds for the Central Universities under Grant No. B200202175, Key Laboratory of Coastal Disaster and Protection of Ministry of Education, Hohai University (201905), and National Natural Science Foundation of China under Grant No.61702159.

## 6. REFERENCES

[1] Wang S., Kang B., Ma J., et al.: A deep learning algorithm using CT images to screen for Corona Virus Disease (COVID-19). medRxiv.

[2] Xu X., Jiang X., Ma C., et al.: Deep Learning System to Screen Coronavirus Disease 2019 Pneumonia. arXiv:2002.09334.

[3] Ying S., Zheng S., Li L., et al.: Deep learning Enables Accurate Diagnosis of Novel Coronavirus (COVID-19) with CT images. medRxiv.

[4] Chen J., Wu L., Zhang J., et al.: Deep learning-based model for detecting 2019 novel coronavirus pneumonia on high-resolution computed tomography a prospective study. medRxiv.

[5] Gozes O., Frid-Adar M., Greenspan H., et al.: Rapid AI Development Cycle for the Coronavirus (COVID-19) Pandemic Initial Results for Automated Detection & Patient Monitoring using De. arXiv:2003.05037.

[6] Zheng C., Deng X., Fu Q., et al.: Deep Learning-based Detection for COVID-19 from Chest CT using Weak Label. medRxiv.

[7] Sethy P. K., Behera S. K., et al.: Detection of Coronavirus Disease (COVID-19) Based on Deep Features. arxiv: 202003.0300.

[8] Li L., Qin L., Xu Z., et al.: Artificial Intelligence Distinguishes COVID-19 from Community-Acquired Pneumonia on Chest CT. Radiology, 2020.
[9] Jin C., Chen W., Cao Y., et al.: Development and Evaluation of an AI System for COVID-19 Diagnosis. medRxiv.
[10] Shi F., Xia L., Shan F., et al.: Large-Scale Screening of COVID-19 from Community-Acquired Pneumonia using Infection Size-Aware Classification. arXiv:2003.09860.
[11] Jin S., Wang B., Xu H., et al.: AI-assisted CT imaging analysis for COVID-19 screening: Building and deploying a medical AI system in four weeks. medRxiv.
[12] Hemdan E. E. D., Shouman M. A., Karar M. E.: COVIDX-Net A Framework of Deep Learning Classifiers to Diagnose COVID-19 in X-Ray Images. arXiv:2003.11055.
[13] Wang S., Zha Y., Li W., et al.: A Fully Automatic Deep Learning System for COVID-19 Diagnostic and Prognostic Analysis. medRxiv
[14] Apostolopoulos I. D., Bessiana T.: COVID-19: Automatic detection from X-Ray images utilizing transfer learning with convolutional neural networks. arXiv: 2003.11617.
[15] Asmaa A., Mohammed M. A., Mohamed M. G.: Classification of COVID-19 in chest X-ray images using DeTraC deep convolutional neural network. arXiv.
[16] Zhang J., Xie Y., Li Y., et al.: COVID-19 Screening on Chest X-ray Images Using Deep Learning based Anomaly Detection. arXiv:2003.12338.
[17] Biraja G., Allan T.: Estimating Uncertainty and Interpretability in Deep Learning for Coronavirus (COVID-19) Detection. arXiv:2003.10769.
[18] Fu M., Yi S., Zeng Y., et al.: Deep Learning-Based Recognizing COVID-19 and other Common Infectious Diseases of the Lung by Chest CT Scan Images. medRxiv.
[19] Muhammad E. H., Tawsifur R., Amith K. et al.: Can AI help in screening viral and COVID‐19 pneumonia? arXiv:2003.13145.
[20] Ali N., Ceren K., Ziynet P.: Automatic detection of coronavirus disease (COVID-19) using X-ray images and deep convolutional neural networks. arXiv:2003.13145.
[21] Lamia N. M., Kadry A. E., Haytham H. E., et al.: Automatic X-ray COVID-19 Lung Image Classification System based on Multi-Level Thresholding and Support Vector Machine. medRxiv.
[22] Wang L., Alexander W.: COVID-Net A Tailored Deep Convolutional Neural Network Design for Detection of COVID-19 Cases from Chest Radiography Images arXiv:2003.09871.
[23] Halgurd S. M., Aras T. A., Kayhan Z. G., et al.: Diagnosing COVID-19 Pneumonia from X-Ray and CT Images using Deep Learning and Transfer Learning Algorithms. arXiv:2004.00038.
[24] Muhammad F., Abdul H.: COVID-ResNet A Deep Learning Framework for Screening of COVID19 from Radiographs. arXiv:2003.14395.
[25] Nour E. M. K., Mohamed H. N. T., Aboul E. H., et al.: Detection of Coronavirus (COVID-19) Associated Pneumonia based on Generative Adversarial Networks and a Fine-Tuned Deep Transfer Learning Model using Chest X-ray Dataset. arXiv:2004.01184.
[26] Parnian A., Shahin H., Farnoosh N., et al.: COVID-CAPS A CAPSULE NETWORK-BASED FRAMEWORK FOR IDENTIFICATION OF COVID-19 CASES FROM X-RAY IMAGES. arXiv:2004.02696.
[27] Ioannis D. A., Aznaouridis I. S., Tzani A. M., et al.: Extracting possibly representative COVID-19 Biomarkers from X-Ray images with Deep Learning approach and image data related to Pulmonary. arXiv:2004.00338.
[28] Gozes O., Frid-Adar M., Nimrod S., et al.: Coronavirus Detection and Analysis on Chest CT with Deep Learning. arXiv:2004.02640.
[29] Umut O., Saban O., Mucahid B.: Coronavirus (COVID-19) Classification using Deep Features Fusion and Ranking Technique. arXiv:2004.03698.
[30] Dalia E., Aboul H., Hassan A. E.: GSA-DenseNet121-COVID-19 a Hybrid Deep Learning Architecture for the Diagnosis of COVID-19 Disease based on Gravitational Search Optimization Algorithm. arXiv:2004.05084.
[31] Md. R. K., Till D., Dietrich R. S., et al.: DeepCOVID-Explainer: Explainable COVID-19 Predictions Based on Chest X-ray Images. arXiv:2004.04582.
[32] Md Z. A., M M S. R., Mst S. N., et al.: COVID-MTNet COVID-19 Detection with Multi-Task Deep Learning Approaches. arXiv:2004.03747.
[33] Eduardo J. S. L., Pedro L. S., Rodrigo S., et al.: Towards an Efficient Deep Learning Model for COVID-19 Patterns Detection in X-ray Images. arXiv:2004.05717.
[34] Oh Y., Park S., Ye J. C.: Deep Learning COVID-19 Features on CXR using Limited Training Data Sets. arXiv:2004.05758.
[35] Lawrence O. H., Rahul P., Dmitry B., et al.: Finding COVID-19 from Chest X-rays using Deep Learning on a Small Dataset. arXiv:2004.02060.
[36] Li X., Li C., Zhu D.: COVID-MOBILEXPERT ON-DEVICE COVID-19 SCREENING USING SNAPSHOTS OF CHEST X-RAY. arXiv:2004.03042.
[37] Asif I. K., Junaid L. S., Mudasir B., et al.: CoroNet: A Deep Neural Network for Detection and Diagnosis of COVID-19 from Chest X-ray Images. arXiv:2004.04931.
[38] Hu S., Gao Y., Niu Z., et al.: Weakly Supervised Deep Learning for COVID-19 Infection Detection and Classification from CT Images. arXiv:2004.06689.
[39] Rodolfo P., Diego B., Lucas O. T., et al.: COVID-19 IDENTIFICATION IN CHEST X-RAY IMAGES ON FLAT AND HIERARCHICAL CLASSIFICATION SCENARIOS. arXiv:2004.05835.
[40] Arvan M., Pietro A. C., Samira Z., et al.: Radiologist-Level COVID-19 Detection Using CT Scans with Detail-Oriented Capsule Networks. arXiv:2004.07407.
[41] Sivaramakrishnan R., Jen S., Philip O. A., et al.: Iteratively Pruned Deep Learning Ensembles for COVID-19 Detection in Chest X-rays. arXiv:2004.08379.
[42] Mohammad R., Abolfazl A.: A New Modified Deep Convolutional Neural Network for Detecting COVID-19 from X-ray Images. arXiv:2004.08052.
[43] Shervin M., Rahele K., Milan S., et al.: Deep-COVID: Predicting COVID-19 From Chest X-Ray Images Using Deep Transfer Learning. arXiv:2004.09363.
[44] Sara H. K., Peyman H. K., Michal J. W., et al.: Automatic Detection of Coronavirus Disease (COVID-19) in X-ray and CT Images: A Machine Learning-Based Approach. arXiv:2004.10641.
[45] Narinder S. P., Sonali A.: Automated diagnosis of COVID-19 with limited posteroanterior chest X-ray images using fine-tuned deep neural networks. arXiv:2004.11676.
[46] Matteo P., Luigi C., Giuseppe P.: A Light CNN for detecting COVID-19 from CT scans of the chest. arXiv:2004.12837.
[47] Sanhita B., Sushmita M., Nilanjan S.: Deep Learning for Screening COVID-19 using Chest X-Ray Images. arXiv:2004.10507.

[48] Li T., Han Z., Wei B., et al.: Robust Screening of COVID-19 from Chest X-ray via Discriminative Cost-Sensitive Learning. arXiv:2004.12592.
[49] Zhang Y., Niu S., Qiu Z., et al: COVID-DA: Deep Domain Adaptation from Typical Pneumonia to COVID-19. arXiv:2005.01577.
[50] Brian D. G., Corey J., Can Z., et al.: Intra-model Variability in COVID-19 Classification Using Chest X-ray Images. arXiv:2005.02167.
[51] Yeh C., Cheng H., Wei A.: A Cascaded Learning Strategy for Robust COVID-19 Pneumonia Chest X-Ray Screening. arXiv:2004.12786.
[52] Lv D., Qi W., Li Y.: A cascade network for Detecting COVID-19 using chest x-rays. arXiv:2005.01468.
[53] Sun L., Mo Z, Yan F., et al.: Adaptive Feature Selection Guided Deep Forest for COVID-19 Classification with Chest CT. arXiv:2005.03264.
[54] Mehmet Y., Mete A., Aysen D., et al.: Convolutional Sparse Support Estimator Based Covid-19 Recognition from X-ray Images. arXiv:2005.04014.
[55] Zhou M., Chen Y., Yang D., et al.: Improved deep learning model for differentiating novel coronavirus pneumonia. medRxiv.
[56] Chen X., Yao L., Zhang Y.: Residual Attention U-Net for Automated Multi-Class Segmentation of COVID-19 Chest CT Images. arXiv:2004.05645.
[57] Zhou T., Canu S., Ruan S.: An automatic COVID-19 CT segmentation based on U-Net with attention mechanism. arXiv:2004.06673.
[58] Yu Q., Yun L., Jing X.: MiniSeg: An Extremely Minimum Network for Efficient COVID-19 Segmentation. arXiv:2004.09750.
[59] Fan D., Zhou T., Ji G., et al.: Inf-Net: Automatic COVID-19 Lung Infection Segmentation from CT Scans. arXiv:2004.14133.
[60] Wu Y., Gao S., Mei J., et al.: JCS: An Explainable COVID-19 Diagnosis System by Joint Classification and Segmentation. arXiv:2004.07054.
[61] Xi O., Jiayu H., Liming X., et al.: Dual-Sampling Attention Network for Diagnosis of COVID-19 from Community-Acquired Pneumonia. arXiv:2005.02690.
[62] Tahereh J., Morteza H., Zohreh A., et al.: CovidCTNet: An Open-Source Deep Learning Approach to Identify Covid-19 Using CT Image. arXiv:2005.03059.
[63] Kang H., Xia L, Yan F., et al.: Diagnosis of Coronavirus Disease 2019 (COVID-19) with Structured Latent Multi-View Representation Learning. arXiv:2005.03227.
[64] Mehmet Y., Mete A., Aysen D., et al.: Convolutional Sparse Support Estimator Based Covid-19 Recognition from X-ray Images. arXiv:2005.04014.
[65] Sampa M., Seungwan J., Seiyon L., et al.: Multi-Channel Transfer Learning of Chest X-ray Images for Screening of COVID-19. arXiv:2005.05576.
[66] Zhou J., Jing B., Wang Z. et al.: SODA: Detecting Covid-19 in Chest X-rays with Semi-supervised Open Set Domain Adaptation. arXiv:2005.11003.
[67] Mizuho N., Shunjiro N., Hidetoshi M., et al.: Automatic classification between COVID-19 pneumonia, non-COVID-19 pneumonia, and the healthy on chest X-ray image: combination of data augmentation methods. arXiv:2006.00730.
[68] Soumick C., Fatima S., Chompunuch S., et al.: Exploration of Interpretability Techniques for Deep COVID-19 Classification using Chest X-ray Images. arXiv:2006.02570.
[69] Germán G., Aurelia B., José M., et al.: UMLS-ChestNet: A deep convolutional neural network for radiological findings, differential diagnoses and localizations of COVID-19 in chest x-rays. arXiv:2006.05274.
[70] Morteza H., Seyedehnafiseh M., Abolfazl Z., et al.: Improving performance of CNN to predict likelihood of COVID-19 using chest X-ray images with preprocessing algorithms. arXiv:2006.12229.
[71] Chen X., Yao L, Zhou T., et al.: Momentum Contrastive Learning for Few-Shot COVID-19 Diagnosis from Chest CT Images. arXiv:2006.13276.
[72] Manu Siddhartha, Avik Santra: COVIDLite: A depth-wise separable deep neural network with white balance and CLAHE for detection of COVID-19. arXiv:2006.13873.
[73] Mahesh G., Sweta J.: Stacked Convolutional Neural Network for Diagnosis of COVID-19 Disease from X-ray Images. arXiv:2006.13817.
[74] Rohit L., Ashrika G., Viraj K., et al.: Automated Detection of COVID-19 from CT Scans Using Convolutional Neural Networks. arXiv:2006.13212.
[75] Abdolkarim S., Maryam S., Arash M. et al.: A Novel and Reliable Deep Learning Web-Based Tool to Detect COVID-19 Infection from Chest CT-Scan. arXiv:2006.14419.
[76] Zhong Y. et al.: Using Deep Convolutional Neural Networks to Diagnose COVID-19 From Chest X-Ray Images. arXiv:2007.09695.
[77] Md. K., Md. A., Md. T., et al.: CVR-Net: A deep convolutional neural network for coronavirus recognition from chest radiography images. arXiv:2007.11993.
[78] Imon B., Priyanshu S., Saptarshi P., et al.: Was there COVID-19 back in 2012? Challenge for AI in Diagnosis with Similar Indications. arXiv:2006.13262.
[79] Wynants L., Van Calster B., Bonten M. M. J., et al.: Systematic review and critical appraisal of prediction models for diagnosis and prognosis of COVID-19 infection. medRxiv.
[80] Bullock J., Pham K. H., Lam C. S. N., et al.: Mapping the landscape of artificial intelligence applications against COVID-19. arXiv:2003.11336.
[81] Nguyen T. T.: Artificial Intelligence in the Battle against Coronavirus (COVID-19): A Survey and Future Research Directions.
[82] Shi F., Wang J., Shi J., et al.: Review of artificial intelligence techniques in imaging data acquisition, segmentation and diagnosis for covid-19. IEEE Reviews in Biomedical Engineering, 2020.
[83] Ilyas M., Rehman H., Naitali A.: Detection of Covid-19 From Chest X-ray Images Using Artificial Intelligence: An Early Review. arXiv:2004.05436.
[84] Anwaar U., Asim K., Douglas G., et al.: COMPUTER VISION FOR COVID-19 CONTROL A SURVEY. arXiv:2004.09420.
[85] Ying M., Susiyan J., Daniel N., et al.: Data-driven Analytical Models of COVID-2019 for Epidemic Prediction, Clinical Diagnosis, Policy Effectiveness and Contact Tracing: A Survey. arXiv:2006.13994.
[86] Chen J., Li K., Zhang Z., et al.: A Survey on Applications of Artificial Intelligence in Fighting Against COVID-19. arXiv:2007.02202.
[87] Afshin S.i, Marjane K., Roohallah A., et al.: Automated Detection and Forecasting of COVID-19 using Deep Learning Techniques: A Review. arXiv:2007.10